\documentclass[lettersize,journal]{IEEEtran}
\usepackage{amsmath,amsfonts}
\usepackage{algorithmic}
\usepackage{algorithm}
\usepackage{array}
\usepackage[caption=false,font=normalsize,labelfont=sf,textfont=sf]{subfig}
\usepackage{textcomp}
\usepackage{stfloats}
\usepackage{url}
\usepackage{verbatim}
\usepackage{graphicx}
\usepackage{cite}
\usepackage{xcolor}
\usepackage{caption}
\usepackage{subcaption}
\usepackage{pgfplots}
\pgfplotsset{width=10cm,compat=1.9}
\usepackage[
    colorlinks=false,
    pdfborder={0 0 0}
]{hyperref}
\usepackage{soul}

\usepackage{fancyhdr}

\begin{document}

\title{IUP: Integrated and Programmable User Plane for Next-Generation Mobile Networks}
\author{
    \IEEEauthorblockN{
    Chieh-Chun Chen\IEEEauthorrefmark{1},
    Chia-Yu Chang\IEEEauthorrefmark{2}, Navid Nikaein\IEEEauthorrefmark{1}}\\
    \IEEEauthorblockA{\IEEEauthorrefmark{1}EURECOM, Sophia-Antipolis, France, email:\{chieh-chun.chen, navid.nikaein\}@eurecom.fr}\\
    \IEEEauthorblockA{\IEEEauthorrefmark{2}Nokia Bell Labs, Antwerp, Belgium, email: chia-yu.chang@nokia-bell-labs.com}

}
\maketitle

\thispagestyle{fancy}
\fancyhf{}
\fancyhead[L]{\small
    This manuscript has been accepted for publication on IEEE Network Magazine.\\
    © 2025 IEEE. Personal use of this material is permitted.  Permission from IEEE must be obtained for all other uses, in any current or future media, including reprinting/republishing this material for advertising or promotional purposes, creating new collective works, for resale or redistribution to servers or lists, or reuse of any copyrighted component of this work in other works.
}
\renewcommand{\headrulewidth}{0pt}

\begin{abstract}
Mobile networks evolve on a regular basis to meet the requirements of a rapidly changing application ecosystem; hence, a future-proof design is key to getting the most out of their lifecycle. In comparison to other access networks, one major issue with the 5G Radio Access Network (RAN) is that it behaves as a ``fat Layer 2" entity, resulting in disparities in Internet Protocol (IP) flow traffic control and radio resource allocation. In this article, we propose an innovative design - Integrated User Plane (IUP) - that incorporates User Plane Function (UPF) functionalities into RAN, and we introduce the Integrated Data Flow Control (IDFC) sublayer with a new traffic management pipeline and various programmable rules. To understand its implications for crucial mobility user cases, a detailed analysis of how IUP interacts with Control Plane (CP) network functions is conducted. Finally, our IUP prototype shows benefits including a 50\% saving in both latency and overhead, converged IUP and non-Third-Generation Partnership Project (3GPP) networks for seamless connectivity, and real-time UP programmability in both traffic control and resource allocation via the O-RAN framework.

\end{abstract}


\section{Introduction} \label{sec:intro}

The Fifth Generation (5G) system is designed with evolving principles, regularly upgraded with each generation to support not only the Third-Generation Partnership Project (3GPP) services but also IP-based services, such as web browsing and other forms of Internet access~\cite{TS23-501}. 
However, with the rapid growth of the application ecosystem and the rise of over-the-top services, application traffic has become the primary source of network usage for end-users~\cite{5g-app-challenges}.
Unlike the relatively stable evolution of mobile network infrastructure, application traffic characteristics often change faster due to frequent software updates and new releases, outpacing the typical ten-year mobile network upgrade cycle.

In this context, we envision that next-generation mobile networks adopting an application-centric design, featuring a flattened and routing/switching-based structure to enable a converged network~\cite{jamshed2024nonterrestrialnetworks6gintegrated}. This new focus will enhance interconnections between points of presence and provide real-time data for application decision-making. From an application perspective, it can interface with access network technologies (e.g., Wi-Fi) and network infrastructure (e.g., router), allowing applications to optimize traffic delivery path(s) and efficiency.



As of today, achieving this goal in 5G faces three main challenges: (1) Separating Radio Access Network (RAN) functions — whether splitting the Distributed Unit (DU) and Centralized Unit (CU) or dividing the User Plane (UP) from the Control Plane (CP) — can cause inconsistencies between traffic control and radio resource allocation; (2) RAN handles aggregated service data flows through Data Radio Bearers (DRBs), limiting its ability to provide detailed link-layer insights for each traffic flow; (3) tunneling and encapsulation between the RAN and User Plane Function (UPF) introduce processing overhead, increase latency, and hinder direct IP routing for local User Equipment (UE) communications.

Beyond the challenges in 5G UP, enabling RAN programmability follows different approaches. Architecturally, the O-RAN framework~\cite{understanding-o-ran} enables dynamic RAN control through xApps on the Near-Real Time RAN Intelligent Controller (NearRT-RIC), handling tasks like radio resource allocation~\cite{flexslice}. From an optimization perspective, cross-layer coordination improves traffic management, addressing issues like buffer bloat~\cite{TC-RAN} and prioritizing short flows~\cite{outRAN}. However, achieving a fully programmable RAN that adapts to real-time network conditions requires integrating both architectural control and optimization techniques.

\begin{figure*}[!t]
  \centering
  \includegraphics[width=\textwidth]{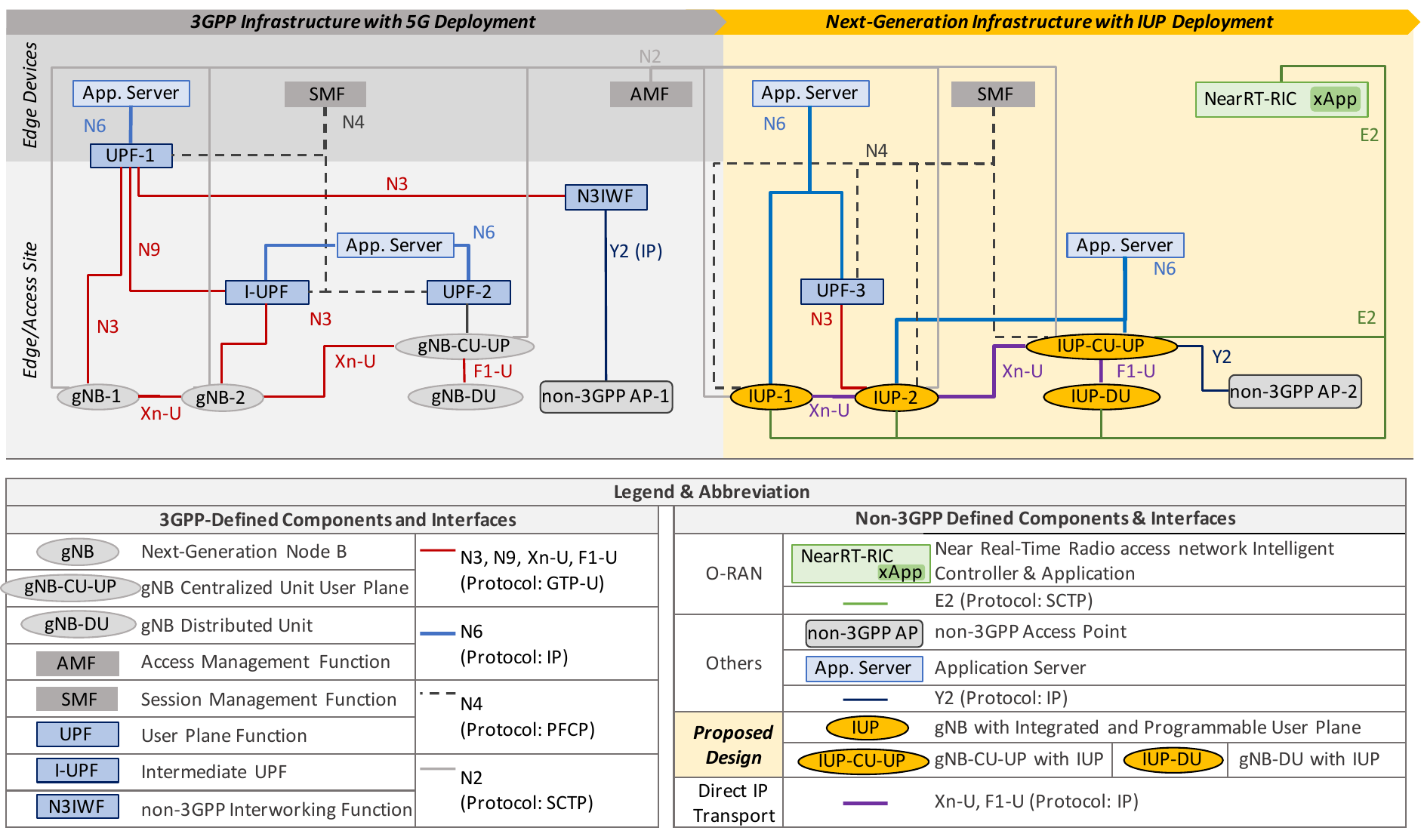}
  \caption{From 5G disaggregated user plane deployment to IUP in next-generation mobile networks.}
  \label{fig:intro}
\end{figure*}

\textbf{Contributions.}
This article aims to integrate UP across RAN and UPF into a unified entity known as the Integrated User Plane (IUP), envisioned as the out-of-the-box deployment for next-generation mobile networks, as shown in Fig.~\ref{fig:intro}.
IUP simplifies network infrastructures by reducing protocol overhead, e.g., GPRS Tunneling Protocol User Plane (GTP-U), and enhances UP programmability from IP flows to radio resources, while also enabling convergence with non-3GPP networks.
Finally, we evaluate its performance in an over-the-air testbed and show its benefits among several aspect.

\section{Implications and Benefits} \label{sec:challenges}
Based on the above vision provided by IUP, below we identify three key benefits and potential improvements.

\subsection{Reduced Protocol Overhead, Processing, and Latency}
\label{sec:challenges:latency}

5G RAN UP includes not only sublayers for air-interface communication, but also functions like GTP-U processing for tunneling to UPF over N3 interface.
In 5G, as shown in left part of Fig.~\ref{fig:intro}, centralized UPF and distributed RAN (e.g., gNB-1) are commonly used, resulting in a "long N3".
However, advances in RAN cloudification and mobile edge computing~\cite{TS23-558}, enables a distributed UPF that shortens N3 link by deploying UPF or Intermediate UPF (I-UPF)~\cite{TS23-501} along with application servers at the edge (e.g., gNB-2 and gNB-CU-UP).

In contrast, as shown in right part of Fig.\ref{fig:intro}, the proposed IUP\footnote{IUP also supports deployment with an extra hop to UPF and application server, serving as an intermediate stage of deployment (e.g., IUP-2 in Fig.\ref{fig:intro}) to ease network upgrades without requiring overall network renewal.} removes the need for the N3 interface (e.g., IUP-1 and IUP-CU-UP) and the corresponding GTP-U processing.
This also avoids IP fragmentation caused by different Maximum Transmission Unit (MTU) sizes between GTP Protocol Data Unit (PDU) and IP packet~\cite{data-plane-fragmentation}.

Regarding latency reduction, while solutions like network slicing have already shown their capability to meet the hierarchical Quality of Service (QoS) requirements specified in service level agreements ~\cite{TS28-530}, IUP operates independently of network slicing and reduces latency from two sources.
First, by minimizing backhaul transmission delays between centralized/distributed UPF and gNB, IUP better manages the packet delay budget~\cite{TS23-501}, not only from a RAN perspective but also across backhaul network.
Second, it bypasses N3-related processing and avoids encapsulation overhead.

\subsection{Programmable User Plane \& Simplified Control Plane}
\label{sec:challenges:programmable}
Existing UP programmability solutions focus on two approaches: 
(1) introducing programmability to the packet processing pipeline in the UPF~\cite{TS29-244}, such as P4 on programmable hardware platform~\cite{P4-based-UPF};
and (2) leveraging O-RAN framework to develop xApps on NearRT-RIC for dynamically controlling RAN functions~\cite{flexslice}.
As mentioned before, these two approaches must be coordinated to avoid conflicts between control decisions.
Thanks to the integration of UPF into RAN, IUP facilitates a coordinated approach to control both IP packet processing and lower-layer processing of the mapped DRBs.


Moreover, IUP is designed to be compatible with existing CP functions of the CN, as illustrated in Fig.~\ref{fig:intro}.
It retains Packet Forwarding Control Protocol (PFCP) functionalities, allowing it to interact with Session Management Function (SMF) via the N4 interface and reuse PDU session management procedures. Also, the N2 interface toward Access and Mobility Management Function (AMF) is preserved for connection management.
As IUP becomes a single point for data delivery, it can avoid extra control messages and data forwarding in some mobility scenarios (cf. Section~\ref{sec:use-cases}).

\subsection{Seamlessly Converged Network}
\label{sec:challenges:converged}
Network convergence with non-3GPP networks can be achieved in different ways.
One is to design a compatibility layer or network function, such as Non-3GPP Interworking Function (N3IWF)~\cite{TS24-502}, to connect 5G and non-3GPP networks.
Alternatively, IUP takes another approach by incorporates IP into RAN, enabling Internet-wide routing and providing interoperability regardless of underlying link technologies.
Therefore, it allows real-time applications aware of latency across multiple paths to destination, facilitating different traffic flows (e.g., extended reality objects) using individual routes.
As shown in Fig.~\ref{fig:intro}, IUP and non-3GPP AP-2 can be converged seamlessly without the need of an extra gateway in between.

\subsection{Summary}
First, the proposed IUP reduces N3-related overhead and processing while also simplifying the end-to-end data delivery path in the mobile network (cf. Section~\ref{sec:results:latency}).
Additionally, IUP enhances UP programmability by extending control from the IP layer (and above) to the radio link layer, providing a unified framework for managing both packet flows and radio resources (cf. Section~\ref{sec:results:programmable}).
Finally, IUP enables universal connectivity by acting as a Layer 3 device, seamlessly integrating diverse access technologies via the IP protocol (cf. Section~\ref{sec:results:converged}).

\begin{figure*}[!t]
  \centering
  \includegraphics[width=\textwidth]{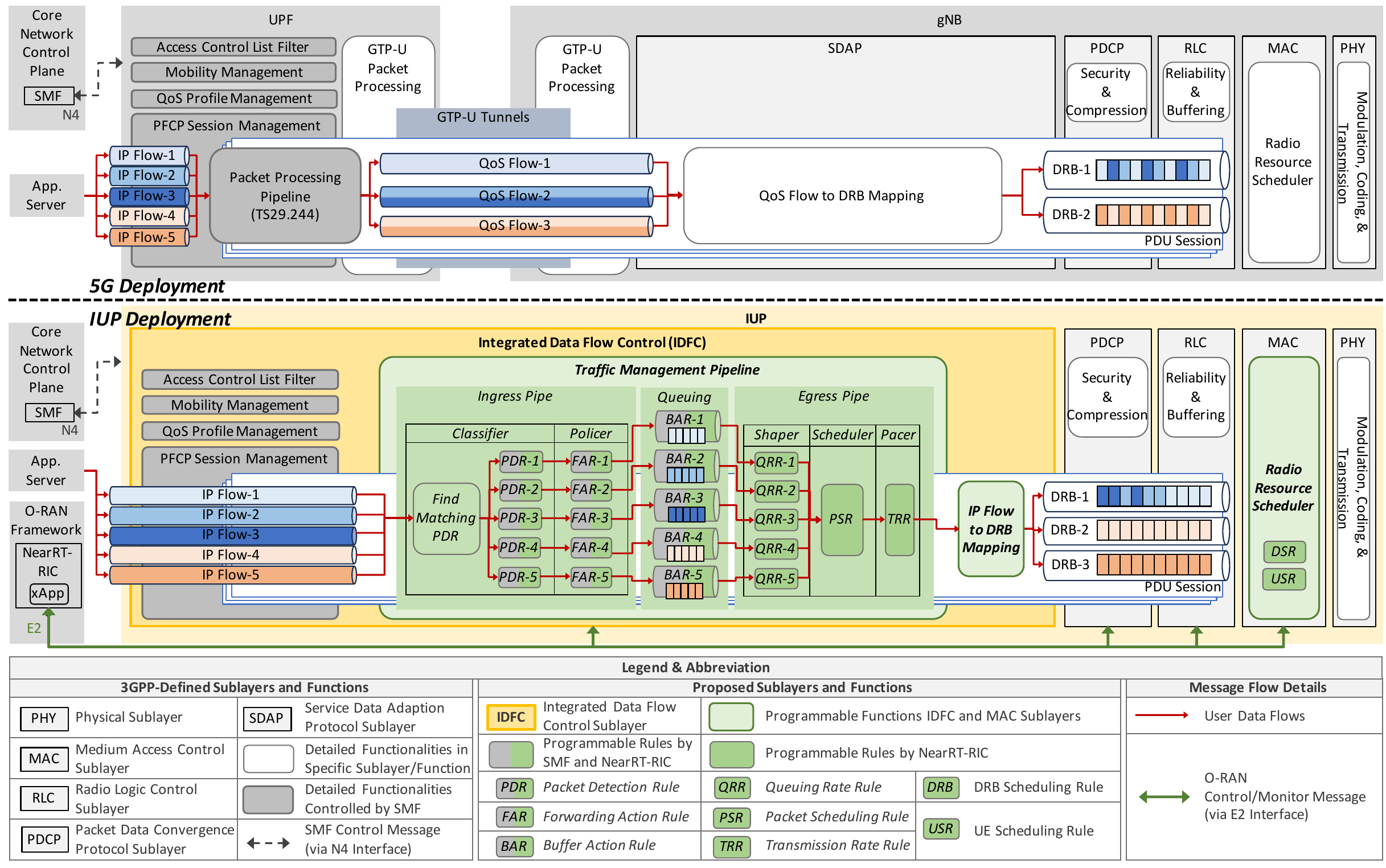}
  \caption{Architecture for 5G and IUP deployments: user data flows, programmable functions and rules, and control plan interfaces.}
  \label{fig:arch}
\end{figure*}

\section{IUP Architecture}
\label{sec:architecture}
In the following, we elaborate on the architecture of IUP, as shown in the lower half of Fig.~\ref{fig:arch}.
Meanwhile, the upper half of Fig.~\ref{fig:arch} depicts the 5G deployment.

\subsection{Integrated User Plane}
The proposed architecture introduces an Integrated Data Flow Control (IDFC) sublayer\footnote{The IDFC sublayer replaces the current SDAP sublayer and is not backward compatible; however, such compatibility can be maintained by deploying extra GTP-U and SDAP processing in IUP (cf. Section~\ref{subsec:4-d}).} that not only consolidates UPF functionalities into RAN but also extends the 3GPP-defined packet processing pipeline~\cite{TS29-244} to allow for programmability through the whole UP, i.e., from IP traffic control to radio resource allocation.
In particular, Fig.~\ref{fig:arch} depicts how user data flows are delivered from the application server, via several RAN sublayers, and finally to end-users.

\textbf{5G Deployment.}
As seen in the upper half of Fig.~\ref{fig:arch}, user data is transmitted over the PDU session, and the context of PDU session is addressed by PFCP session management to determine how to process packets within the UPF pipeline.
In details, SMF can control this pipeline by defining rules, e.g., QoS enforcement rules, to decide how UPF translates IP flows into QoS flows and assigns QoS attributes (e.g., 5QI~\cite{TS23-501}).
Subsequently, QoS flows are sent to gNB via GTP-U tunnels, and the gNB assigns them to DRBs at Service Data Adaptation Protocol (SDAP) sublayer.
Finally, the remaining processing is performed by gNB, such as packet compression, buffering, and radio resource allocation.
Note that while SMF can control the packet processing pipeline\footnote{This can further rely on the rules and charging policies provided by Policy Control Function (PCF).}, it does not have visibility of lower-layer information.
On the other hand, the Medium Access Control (MAC) radio resource scheduler lacks insight into each application flow because multiple IP flows are aggregated into a QoS flow, and multiple QoS flows are mapped into a DRB (e.g., IP-flow-2 and IP-flow-3 are aggregated into QoS-flow-2, and QoS-flow-1 and QoS-flow-2 are mapped into DRB-1). Therefore, the UP is not properly integrated for unified programmability.


\vspace{5mm}
\textbf{IUP Deployment.} 
In contrast, as shown in the lower half of Fig.~\ref{fig:arch}, IUP integrates UPF into RAN as the IDFC sublayer while maintaining certain UPF functionalities, such as PDU session establishment, modification, and release via PFCP session management. Therefore, similar to 5G deployment, user data is still sent over PDU sessions. Moreover, the IDFC sublayer can directly handle IP flows, provide granular traffic control in the traffic management pipeline, and map them into DRBs. This pipeline has three main stages with several programmable rules: (1) Ingress pipe (classifier and policer), (2) Queuing, and (3) Egress pipe (shaper, scheduler, and pacer), which will be discussed in the next paragraph. In short, IUP eliminates the need for extra GTP-U processing and intermediate QoS flow translation.




\subsection{Traffic Management Pipeline and Programmable Rules}
The traffic management pipeline in IDFC consists of three main stages and can be programmed with six different rules.
First, in the ingress pipe, incoming IP flows are identified by classifiers based on Packet Detection Rules (PDRs) and then forwarded or dropped by policers based on Forwarding Action Rules (FARs).
Afterwards, packets will be buffered in queues, and each queue will be managed according to the Buffer Action Rules (BAR).
Finally, in the egress pipe, the rate of each queue will be controlled by the shaper using Queuing Rate Rules (QRR), the scheduler will use Packet Scheduling Rules (PSR) to determine which flows can be transmitted, and the Transmission Rate Rule (TRR) is used by the pacer to decide how to pace packets out.

In particular, these six rules are elaborated as below:
\begin{itemize}
    \item \textbf{PDRs} identify IP flows using given information, such as five-tuples. Within multiple PDRs, classifier can analyze the packet to find a matching PDR (and destination queue) and then send it to the policer if a matching PDR is found. Otherwise, the packet will be discarded if no default PDR is provided.
    \item \textbf{FARs} are linked to a queue and determine the actions to be taken by policer before entering the queue, e.g., forwarding or dropping. If no FAR is provided, the packet will go directly to the destination queue.
    \item \textbf{BARs} provide queue management rules for each queue, e.g., First-In-First-Out (FIFO) and Controlled Delay (CoDel). And there is a default queue for processing packets that match default PDR.
    \item \textbf{QRRs} are used by shaper to limit the maximum egress rate of the corresponding queue.
    \item \textbf{PSR} provides the approach (e.g., round-robin or priority-based) to schedule packets across multiple queues.
    \item \textbf{TRR} is used by pacer to control inter-packet time to avoid unnecessary queuing or even packet dropping in lower layer, e.g., Radio Link Control (RLC) buffer.
\end{itemize}
All these rules can be programmed under O-RAN framework; however, as PDRs, FARs, and BARs are stem from 3GPP-defined packet processing, SMF retains control of these rules to complete some CP actions (e.g., charging policy).

Subsequently, IP flows are mapped to the corresponding DRBs for radio link transmission.
For example, in Fig.~\ref{fig:arch}, IP-flow-1, IP-flow-2, and IP-flow-3 are mapped to the same DRB because they have similar traffic management rules.
Finally, the MAC radio resource scheduler, which allocates radio resources to DRBs and UEs, can be programmed using DRB Scheduling Rule (DSR) and UE Scheduling Rule (USR), respectively.
To summarize, the whole UP, from IP flows to radio resources, is programmable by using SMF and NearRT-RIC to fulfill a variety of application needs while optimizing resource utilization.

\subsection{Control Plane Interfaces}

There are two CP interfaces for IUP: (1) N4 interface between SMF and IUP for 3GPP control messages to manage PDU sessions, and (2) E2 interface between NearRT-RIC and IUP for O-RAN control messages to manage the aforementioned programmable rules.

As for the first interface, after UE association and authentication, an IUP is selected by SMF to allocate IP addresses during PDU session establishment.
Then, SMF will communicate with the IUP via PFCP functionalities for tasks such as PDU session modification/release and IP anchoring while enforcing policies set by PCF. From the perspective of SMF, IUP acts as a common UPF. Additionally, SMF can choose different IUP instances based on service requirements and send control messages to the source or target IUP to perform mobility management operations.

For the second interface, xApps on NearRT-RIC control both programmable functions and rules within IUP, as mentioned before.
By integrating UPF into RAN, IUP enables decision-making across multiple layers simultaneously, i.e., from traffic management rules of each IP flow to radio resource scheduling of each DRB/UE, and can be simply combined with other schemes, e.g., network slicing.

\section{Use Cases}
\label{sec:use-cases}
Below, four use cases that would occur in IUP deployment are examined: (1) Handover; (2) Roaming; (3) RAN disaggregation and non-3GPP network interworking; and (4) Compatibility with existing UPF.

\begin{figure*}[t]
  \centering
  \includegraphics[width=\textwidth]{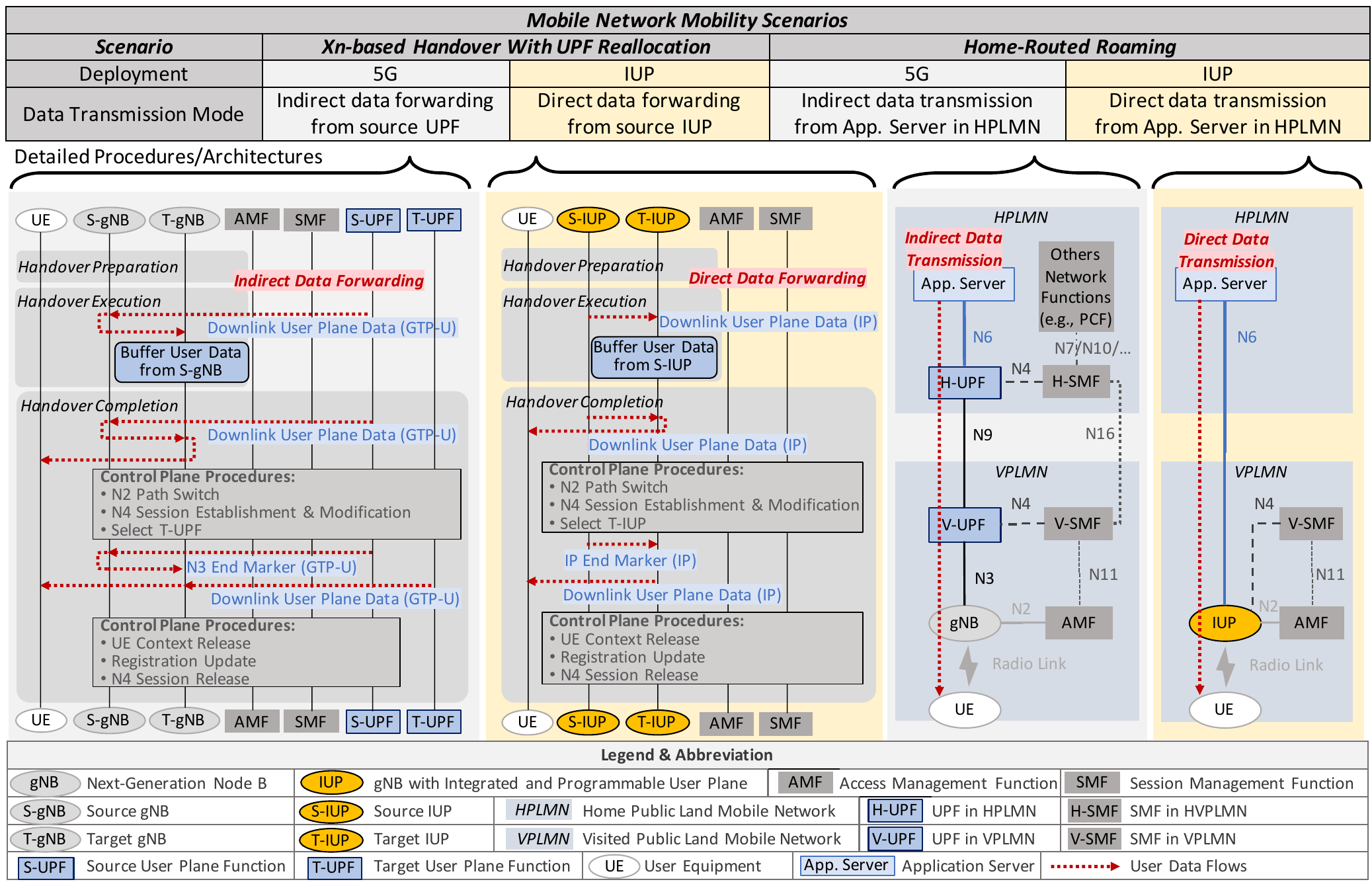}
  \caption{Comparison of 5G and IUP deployments in mobility scenarios: Xn-based handover with UPF reallocation and home-routed roaming.}
  \label{fig:use-case:handover-roaming}
\end{figure*}

\subsection{Handover} \label{subsec:4-a}
The handover process in mobile networks enables the transfer of ongoing data sessions from one cell to another.
However, as the corresponding UPF may be reallocated, both handover processes are analyzed as below, i.e., with or without UPF reallocation\footnote{There are further sub-categories within these two cases, but they have few variations in the UP procedure.}.
In details, the handover process consists of three stages: preparation, execution, and completion, as shown on the left side of Fig.~\ref{fig:use-case:handover-roaming}.

In the case of with UPF reallocation in 5G, the source gNB uses GTP-U tunnels over the Xn-U interface
to forward data from the source UPF to the target gNB.
Such indirect data forwarding will continue until the target UPF is applied as the new anchoring point.
In contrast, the IUP deployment uses a peer-to-peer connection to forward packets since each IUP can act as the anchoring point, allowing IP packets to be forwarded directly between source and target IUPs.
Moreover, because IUP already integrates UPF, there is no scenario without UPF reallocation.
Finally, the CP procedure for IUP handover is similar to the one in~\cite{TS23-502}, relying on N2 path switching done by AMF, as well as PDU session management, IP address allocation, and IUP selection handled by SMF\footnote{The SMF can mitigate service disruptions based on the supported Session and Service Continuity (SSC) mode~\cite{TS23-501}, either preserving the existing IP address or assigning a new one as needed.}.

\subsection{Roaming} \label{subsec:4-b}
In 5G home-routed roaming scenario~\cite{TS23-502}, as shown in the right side of Fig.~\ref{fig:use-case:handover-roaming}, traffic is sent from an application server in Home Public Land Mobile Network (HPLMN) to a UE in Visited PLMN (VPLMN).
User data is traversed through UPF in HPLMN (H-UPF), UPF in VPLMN (V-UPF), gNB of VPLMN, and finally to UE.
This procedure involves extra CP messages between network functions in HPLMN and VPLMN, e.g., SMF in HPLMN (H-SMF) and VPLMN (V-SMF) for session management via the N16 interface.
Also, user data is forwarded between H-UPF and V-UPF over the N9 interface using GTP-U tunnels.
These CP messages and UP forwarding add extra overhead and processing.

Conversely, IUP reduces these overheads, as shown on the right side of Fig.~\ref{fig:use-case:handover-roaming}, where the UE obtains its IP from the V-SMF and communicates directly with the application server in the HPLMN over IUP, bypassing H-UPF and V-UPF as well as eliminating the procedures between H-SMF and V-SMF.
By serving as the default deployment, IUP simplifies traffic routing in roaming scenarios, making home-routed roaming similar to local-breakout roaming, with the only difference being the location of the application server while maintaining the respective traits: low latency for local-breakout roaming and better control for home-routed roaming.

\subsection{RAN Disaggregation and Non-3GPP Interworking} \label{subsec:4-c}
IUP deployment supports RAN disaggregation and interworking with non-3GPP networks.
First, IUP can be disaggregated into IUP-CU-UP and IUP-DU (see Fig.~\ref{fig:intro}), enabling user data flows and downlink data delivery status to be transmitted with an extra IP header, eliminating the need for GTP-U header between gNB-CU-UP and gNB-DU.
Note that the IPSec protocol can be applied for encrypted transportation in between.
Second, IUP can communicate directly with non-3GPP networks (cf. Y2 interface between IUP-CU-UP and non-3GPP AP-2 of Fig.~\ref{fig:intro}).
Not only does this enable interworking without extra network functions, it also empowers real-time control across various access network.

\subsection{Backward Compatibility} \label{subsec:4-d}
To maintain compatibility between IUP and 5G deployments, a key challenge is handling the interface between IUP and the existing UPF.
Therefore, both GTP-U processing and SDAP sublayer are still required during the early deployment.
Specifically, the GTP-U processing ensures that the existing UPF recognizes IUP as an I-UPF, utilizing the N9 interface between UPF and IUP.
While in the event when the existing UPF would view IUP as a gNB via the N3 interface, the SDAP sublayer is needed to map incoming QoS flows into DRBs and the IDFC sublayer will be omitted.
Once existing UPFs are upgraded, the additional GTP-U processing and SDAP sublayer in the IUP can be removed.

\section{Proof-of-Concept Evaluation}
\label{sec:results}
Next, we present the results of our IUP implementation\footnote{Proof-of-Concept implementation is built on top of open-source platforms: OpenAirInterface (OAI) and FlexRIC~\cite{OAIandFlexRIC}.}, showing how IUP can (1) reduce latency and overhead in UP performance, (2) seamlessly converge with non-3GPP network, and (3) provide programmability for IP flow traffic control and radio resource allocation.



\subsection{Latency and Overhead Reduction}
\label{sec:results:latency}
\begin{figure}[t]
    \centering
    \includegraphics[width=0.9\columnwidth]{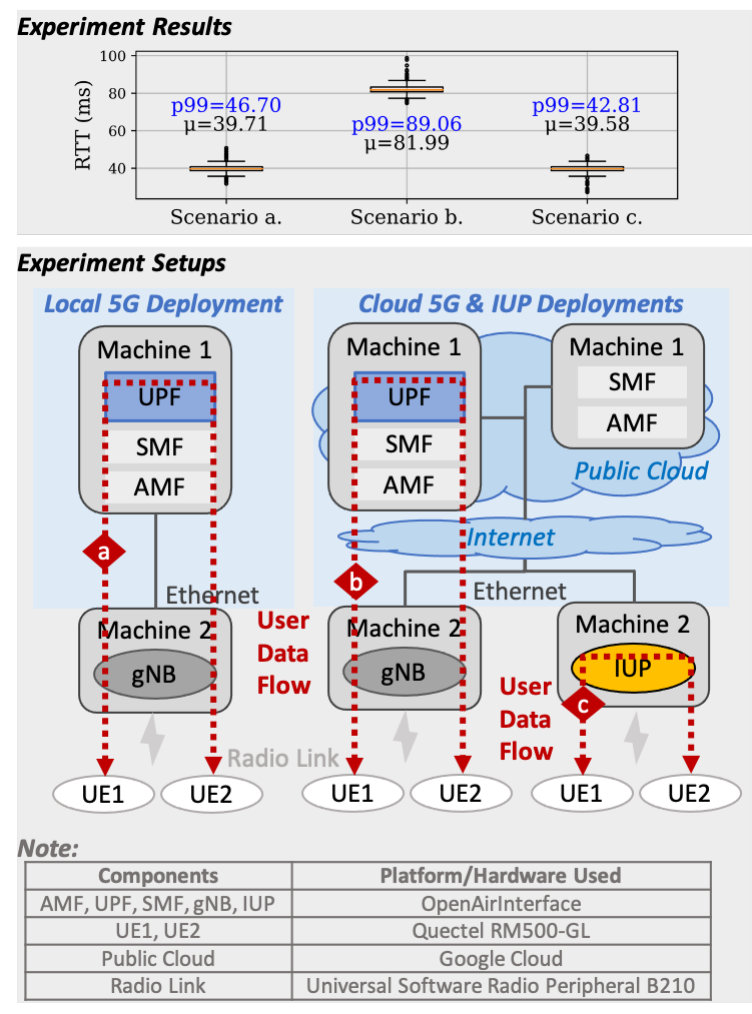}
    \caption{Experiment setup and measured RTT between two UEs in three scenarios: (a) Local 5G, (b) Cloud 5G, and (c) Cloud IUP deployments.}   \label{fig:results:backhaul}
\end{figure}
By integrating UPF and RAN, IUP can reduce latency and data transfer overhead.
As shown in Fig.~\ref{fig:results:backhaul}, we measure the UP performance between two UEs in three scenarios: (a) Scenario a deploys the CN on a server next to the local gNB, creating a ``short N3"; (b) Scenario b places the CN in a public cloud, connecting the local gNB via a ``long N3"; (c) Scenario c deploys part of the CN (excluding UPF) in a public cloud with a local IUP, resulting in ``no N3."

Based on the measured Round-Trip-Time (RTT), the average UP latency is saved by more than 50\% when compared to the cloud deployment, from 81.99\,ms in Scenario b to 39.58\,ms in Scenario c, and is also improved when compared to the local deployment, i.e., Scenario a.
Such improvement is also seen in P99 latency, due to the removal of N3 interface.

Moreover, IUP removes the GTP-U protocol processing and its associated headers in each packet, including outer IP, UDP, and GTP-U headers.
This enables user data to be delivered efficiently.
For instance, when delivering a 60-byte G.729 VoIP packet, the extra GTP header would consume up to 64 bytes, resulting in a total packet size of 124 bytes.
In this case, IUP removes the need for a GTP header, efficiently delivering the 60-byte packet and reducing overhead by 50\%.

\subsection{Converged with Non-3GPP Network}
\label{sec:results:converged}
\begin{figure}[t]
    \centering
    \includegraphics[width=0.9\columnwidth]{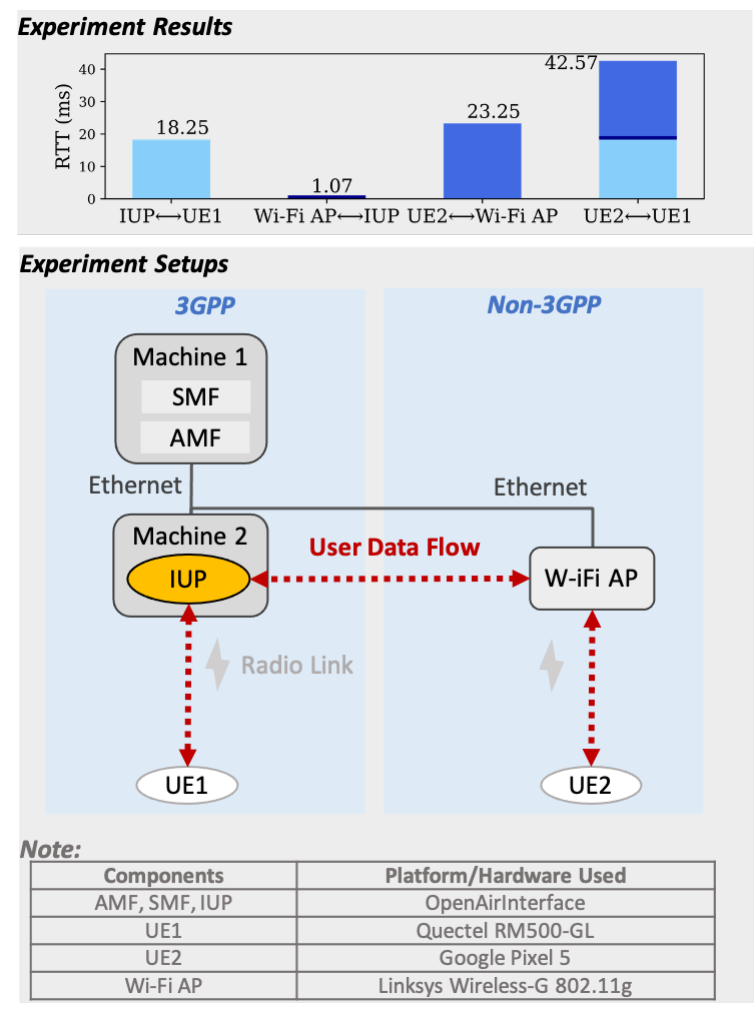}
    \caption{Experiment setup and measured RTT between two UEs served by IUP and WFi AP respectively.}
    \label{fig:results:ubiquitous}
\end{figure}
As shown in Fig.~\ref{fig:results:ubiquitous}, we set up a scenario where IUP serves UE1 and a Wi-Fi Access Point (AP) serves UE2.
This setup demonstrates that IUP can provide connectivity even when one UE is served by a non-3GPP network (compared to Scenario c in Fig.~\ref{fig:results:backhaul}, where IUP serves both UEs).
Thus, IUP works as a Layer 3 router, facilitating direct communication with Wi-Fi AP via IP protocol and shortening the UP path between UE1 and UE2, i.e., avoiding data routing through CN.

Specifically, the average RTT between UE1 and UE2 is approximately 42.57\,ms, as shown in Fig.~\ref{fig:results:ubiquitous}, which is slightly larger than the result in Scenario c of Fig.~\ref{fig:results:backhaul}.
This increase is due to the access delay between Wi-Fi AP and UE2 ($\sim$23.25\,ms), which is larger than the access delay between IUP and UE1 ($\sim$18.25\,ms), while the delay between the IUP and the Wi-Fi AP ($\sim$1.07\,ms) has little impact.

\subsection{Programmability over IP Flows and Radio Resources}
\label{sec:results:programmable}
\begin{figure*}[t]
  \centering
  \includegraphics[width=\textwidth]{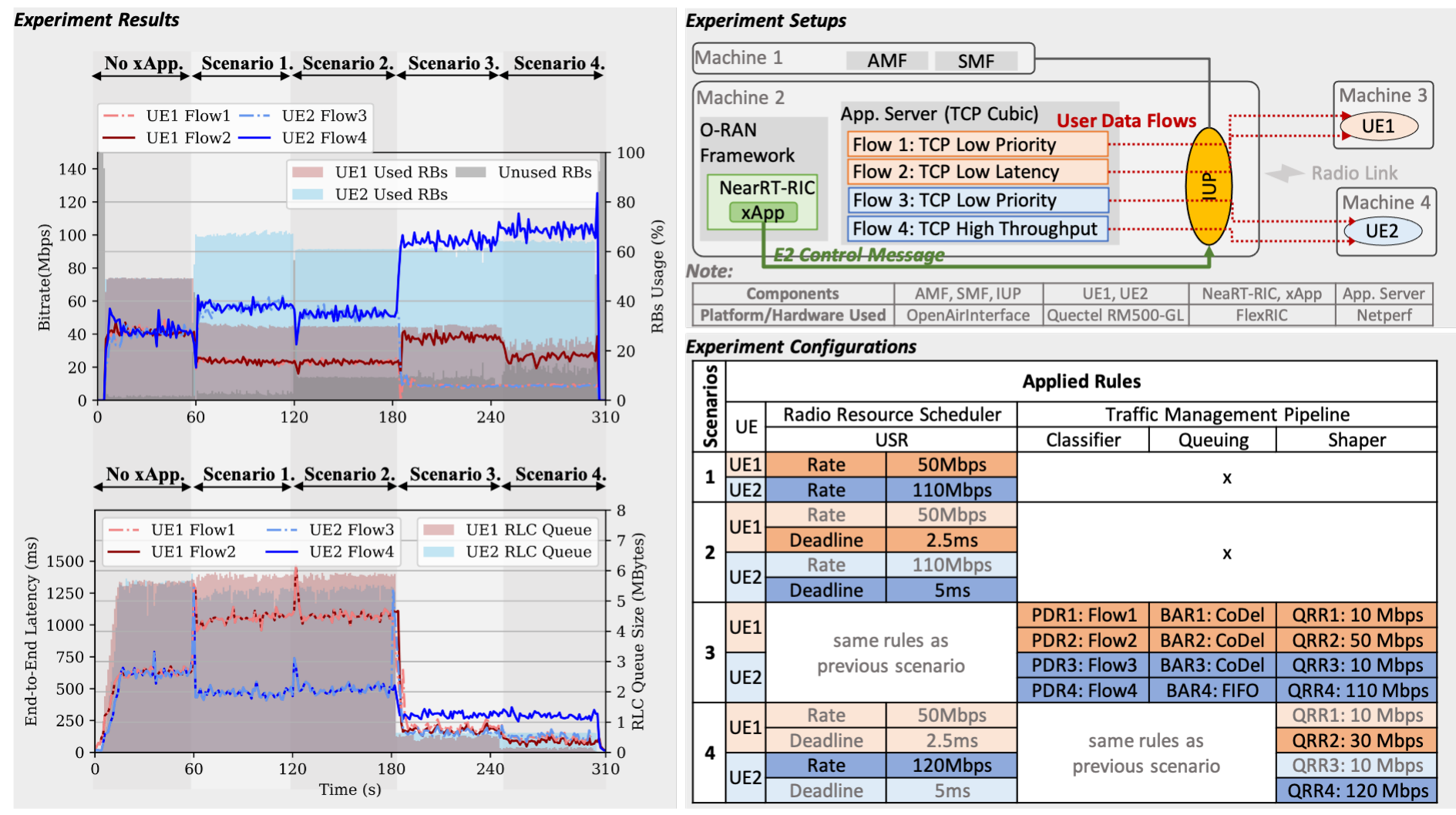}
  \caption{Experiment setup, configurations and results of network and application performance in different scenarios.}
  \label{fig:results:programmability}
\end{figure*}



To show the capability of IUP in terms of programmability, a new xApp is designed to control both IP flows and radio resources, as depicted in Fig.~\ref{fig:results:programmability}.
This xApp enforces rules for the 
traffic management pipeline at the IDFC sublayer and the radio resource scheduler at the MAC sublayer.
For the traffic management pipeline, we modify PDRs, BARs, and QRRs to adjust the classifier, queuing, and shaper, respectively.
For the radio resource scheduler, the scheduling policy is adjusted according to the applied USRs, including the maximum scheduling rate and scheduling deadline.

Moreover, in our experiment setup, two UEs are connected to the IUP\footnote{IUP is configured with 40 MHz, 30 kHz subcarrier spacing, and a TDD frame structure of 7 downlink and 2 uplink slots with single-layer MIMO.}.
Each UE receives two TCP flows with different Differentiated Services Code Point (DSCP) values in the IP header\footnote{The DSCP value CS1 is used for low-priority services, AF11 for low-latency services, and AF21 for high-throughput services.}, which are mapped to the same DRB, i.e., only one DRB per UE.
Then, we modify the rules in four different scenarios and measure network statistics such as Resource Block (RB) usage and RLC queue size, as well as application performance like data rate and end-to-end latency of each flow.



\textbf{Radio Resource Control.}
At the start of our experiment (0\text{–}60\,s in Fig.~\ref{fig:results:programmability}), the xApp is not configured, and a fair share of radio resource is allocated to both UEs to make each flow share the same rate ($\sim$40\,Mbps).
The xApp then applies the USR to the MAC radio resource scheduler at 60\,s (Scenario 1), where radio resources are scheduled proportionately based on the defined maximum scheduling rates.
Thus, each flow of UE1 and UE2 achieves throughput of 25\,Mbps and 55\,Mbps, respectively. 
Note that such proportional scheduling approach does not consider any scheduling deadline; therefore, all available resources are allocated to either UE per time slot, leaving few unused RBs.
In comparison, when scheduling deadlines are introduced in Scenario 2 (120\text{–}180\,s in Fig.~\ref{fig:results:programmability}), radio resources of each slot are allocated accordingly, resulting in more unused RBs (see~\cite{flexslice} for details on the earliest deadline first scheduling algorithm).

Obviously, controlling only the radio resource allocation cannot differentiate application performance,
e.g., both flows per UE have the same data rate and end-to-end latency.
Thus, the traffic management pipeline at the IDFC sublayer is also controlled in the next two scenarios.

\textbf{IP Traffic Control.}
To differentiate IP flows within each DRB, the xApp applies PDRs, BARs, and QRRs along with the previous USR in Scenario 3 (180\text{–}240\,s in Fig.~\ref{fig:results:programmability}).
Specifically, these PDRs classify each flow into individual queues, each of which is controlled by a respective BAR.
For instance, flow 2 uses CoDel active queue management for lower latency\footnote{CoDel is optionally applied to flows 1 and 3 to prevent large delay.}, while flow 4 utilizes a FIFO queue to maximize throughput.
Afterwards, QRRs limit the egress rate of each queue, and a default round-robin scheduler is used in PSR to schedule packets across queues.
As shown in the results from Fig.~\ref{fig:results:programmability}, each flow now behaves differently.
First, due to the applied QRRs, flows 1 and 3 have lower data rates, while flows 2 and 4 have higher data rates.
Additionally, CoDel significantly reduces the end-to-end latency of flow 2 by over 80\%.
Furthermore, the RLC queue size for both UEs decreases largely due to the default bandwidth-delay product pacer~\cite{TC-RAN} applied to TRR, while the RB allocation remains unchanged due to the same USRs.


In the final scenario (240\text{–} 300\,s in Fig.~\ref{fig:results:programmability}), both USRs and QRRs are updated to coordinate traffic control and resource allocation.
We observe that the updated QRR reduces the end-to-end latency of flow 2.
This adjustment lowers the egress rate of flow 2 to ensure that the overall egress rates for UE1 (i.e., QRR1 and QRR2) remain within the maximum scheduling rate specified by the corresponding USR.
Also, both the egress rate of flow 4 and the maximum scheduling rate for UE2 are increased, allowing for more radio resources to be allocated to flow 4.
Note that even more RBs are unused, but we can better satisfy the needs of user data flows, i.e., lower latency for flow 2 and higher throughput for flow 4.



In summary, IP traffic control and radio resource allocation are equally essential to meet diverse needs of applications.
IUP provides programmability from both perspectives, making it a versatile solution for a wide range of use cases.
\section{Discussions and Future Works}
\label{sec:discussion}
This work opens multiple avenues for further exploration.
Despite IUP's advantages, there is still a lack of understanding of the additional processing required to embed traffic management pipelines (or portions of them) within the RAN.
A key concern is scalability, as its per-flow operations may lead to computational overhead. However, unlike conventional centralized UPFs that manage traffic across multiple cells, IUP operates at the per-cell level, significantly reducing scalability demands.
While implementation challenges exist, they can be mitigated through optimized techniques (e.g., tree-based structures) and hardware acceleration (e.g., SmartNICs).


Another open question is a trade-off analysis between the IUP deployment and the current N-to-1 RAN-UPF design is required to verify both cost and energy efficiencies in a scalable IUP deployment.
Additionally, the mapping from IP flows to DRBs is still an open issue, which was done jointly by UPF and RAN in the past, but is now done only on IUP.
As 3GPP and non-3GPP networks converge, it is critical to analyze the present QoS framework and determine how it might be applied to non-3GPP link technology.
Finally, further research is needed to explore how the orchestration and management can enable xApp programmability, allowing the network to automatically adjust rules based on real-time observations to enhance IUP’s adaptability and performance.




\section{Conclusions}
\label{sec:conclusion}


In this article, we introduce a novel concept - IUP - designed for next-generation mobile networks. IUP evolves RAN nodes with integrated UPF functionalities and its programmable UP, extending from IP flows to radio resources.
Additionally, we analyze key use cases to demonstrate how IUP operates in mobility scenarios and assess its compatibility with existing deployments.
Finally, our real-world testbed results highlight several benefits of IUP, including reduced network latency and overhead, seamless convergence between 3GPP and non-3GPP networks, and UP programmability for handling diverse services.

\section*{Acknowledgment}
This work has been funded by the European Union’s Horizon Europe research and innovation program through the project IMAGINE-B5G (G.A no. 101096452) and HORIZON SNS JU ADROIT6G project (G.A no. 101095363) and by the Dutch National Growth Fund 6G flagship project ``Future Network Services''.

\bibliographystyle{IEEEtran}
\bibliography{reference}

\section*{Biographies}

\begin{IEEEbiographynophoto}{Chieh-Chun Chen}
(chieh-chun.chen@eurecom.fr) is a Ph.D. candidate at the Doctoral School of Information Technology, Telecommunications and Electronics at Sorbonne Université, Paris, France. She conducts her research at the Communication Systems department of EURECOM, Biot, France. 
Her research interests include wireless communications and software-defined networking, with a particular focus on network architectures and network slicing for mobile network systems.
\end{IEEEbiographynophoto}

\begin{IEEEbiographynophoto}{Chia-Yu Chang}
(chia-yu.chang@nokia-bell-labs.com) received his Ph.D. degree from Sorbonne Université, France and is currently a Senior Researcher at Nokia Bell Labs, Belgium. He has more than 15 years of experience in algorithm/protocol research on communication systems and real-time applications in both academic and industrial laboratories, including EURECOM Research Institute, MediaTek Communication System Design, Huawei Swedish Research Center, and Nokia Bell Labs. His research interests include wireless communication, computer networking, low-latency low-loss scalable throughput (L4S), and network-application co-design.
\end{IEEEbiographynophoto}

\begin{IEEEbiographynophoto}{Navid Nikaein}
(navid.nikaein@eurecom.fr) is a Professor in the Communication System Department at Eurecom, the CEO of BubbleRAN, a board member of OpenAirInterface Software Alliance. He received the Ph.D. degree in communication systems from the Swiss Federal Institute of Technology, EPFL in 2003. He is leading a research group focusing on experimental system research related to wireless systems, networking, and its intelligent automation. His research contributions are in the areas of software-defined radio access networks, Intelligent connectivity, and real-time  prototypes and salable emulation and simulation for next generation mobile systems.

\end{IEEEbiographynophoto}

\vfill

\end{document}